\documentclass[runningheads,a4paper]{llncs}

\usepackage{amssymb}
\usepackage{graphicx}

\usepackage{url}
\urldef{\mailsa}\path| mahyuddin@usu.ac.id|
\newcommand{\keywords}[1]{\par\addvspace\baselineskip
\noindent\keywordname\enspace\ignorespaces#1}

\begin{document}

\mainmatter  
\title{Extracting Keyword for Disambiguating Name Based on the Overlap Principle
}

\titlerunning{Extracting Keyword for Disambiguating Name Based on ...}
\author{Mahyuddin K. M. Nasution
\thanks{Proceeding of International Conference on Information Technology and Engineering Application (4-th ICIBA), Book 1, 119-125, February 20-21, 2015.}
\authorrunning{M. K. M. Nasution, S. A. M. Noah, S. Saad}
\institute{Information Technology Department, \\ Fakultas Ilmu Komputer dan Teknologi Informasi (Fasilkom-TI)\\
Universitas Sumatera Utara, Padang Bulan, Medan 20155, Sumatera Utara, Indonesia\\
\mailsa\\
}}
\toctitle{}
\tocauthor{}
\maketitle

\begin{abstract} 
Name disambiguation has become one of the main themes in the Semantic Web
agenda. The semantic web is an extension of the current Web in which information
is not only given well-defined meaning, but also has many purposes that contain
the ambiguous naturally or a lot of thing came with the overlap, mainly deals with
the persons name. Therefore, we develop an approach to extract keywords from
web snippet with utilizing the overlap principle, a concept to understand things with
ambiguous, whereby features of person are generated for dealing with the variety of
web, the web is steadily gaining ground in the semantic research.

\end{abstract}

\section{Introduction}
In semantic the disambiguation is the process of identifying related to essence of word, the nature of which is passed on to any object or entity, and also the meaning is embedded to it by how people use it. Basically, the meaning has been stored in the dictionary, the dictionary was based on events that have occurred in the social, and today they have been shared on the web page. The issues of disambiguation therefore is related to the special case of WSD (word sense disambiguation) \cite{schutze1998,mccarthy2004}, especially with the name of someone who is also the words. Today, along with the growth of the web on the Internet, it is difficult to determine a web page associated with the intended person is right and proper, especially with the presence of semantic meaning as a synonymy and polysemy. Therefore, this paper expressed an approach for identifying a person with exploring web snippets.

\section{Research Methodology}
\subsection{Related Work}
Most of works addressed the name disambiguation, among them about preparing information
for person-specific \cite{mann2003}; finding to the association of persons such as the social networks \cite{bekkerman2005}; distinguishing the different persons with keyword/key phrases \cite{li2005}, associating a domain of citations in the scientific papers \cite{han2005}, etc. However, none of the mentioned works attempt to extract dynamic features of person as the current context to do name disambiguation through queries expansion as a way for fighting against explosion of information on the Internet, that is increasing and expanding relationship between the persons and the words continuously, mainly to face the common words like "information", that is an indwell word always for each person in information era.

Semantically, there are motivations of disambiguation problem: 
\begin{enumerate}
\item Meronymy \cite{nguyen2008a}: $x$ is part $y$ or "is-a", part to whole relation - the semantic relation that holds between a part and the whole. In other word, the page for $x$ belong to the categories of $y$. For example, the page for the Barack Obama in Wikipedia\footnote{\url{https://en.wikipedia.org/wiki/Barack_Obama}} belong to the categories 
\begin{enumerate}
\item President of the United State\footnote{\url{https://en.wikipedia.org/wiki/President_of_the_United_States}}, 
\item United States Senate\footnote{\url{https://en.wikipedia.org/wiki/United_States_Senate}}, 
\item Illinois Senate\footnote{\url{https://en.wikipedia.org/wiki/Illinois_Senate_career_of_Barack_Obama}}, 
\item Black people\footnote{\url{https://en.wikipedia.org/wiki/Black_people}}, etc. 
\end{enumerate}
In other case, some entities are associated with multi-categories. For example, Noam Chomsky is a linguist and Noam Chomsky is also a critic of American foreign policy. 
\item Honolomy: $x$ has y as part of itself or "has-a", whole to part relation - the semantic relation that holds between a whole and its parts. For example, in DBLP, the author name "Shahrul Azman Mohd Noah" has a name label as "Shahrul Azman Noah". 
\item Hyponymy \cite{nguyen2008b}: $x$ be subordinate of $y$ or "has-property", subordination - the semantic relation of being subordinate or belong to a lower rank or class. In other word, the page for $x$ has subcategories of $y$. For example, the homepage of "Tengku Mohd Tengku Sembok" has categories pages: Home, Biography, Curriculum Vitae, Gallery, Others, Contact, Links, etc. Some pages also contain name label "Tengku Mohd Tengku Sembok". 
\item Synonymy \cite{lloyd2005,song2007}: $x$ denotes the same as $y$, the semantic relation that holds between two words or can (in the context) express the same meaning. This means that the entity may have multiple name variations/abbreviations in citations across publications. For example, in DBLP, the author name "Tengku Mohd Tengku Sembok" is sometimes written as "T. Mohd T. Sembok", "Tengku M. T. Sembok". \item Polysemy \cite{song2007}: Lexical ambiguity, individual word or phrase or label that can be used to express two or more different meanings. This means that different entities may share the same name label in multiple citations. For example, both "Guangyu Chen" and "Guilin Chen" are used as "G. Chen" in their citations.
\end{enumerate}

\subsection{A Model}
Let $A = \{a_i|i = 1, . . .,M\}$ is a set of person (real-world entities). There is $A_d = \{b_j|j = 1, . . .,N\}$ as a set of ambiguous names which need to be disambiguated, e.g., $\{$"John Barnes",$\dots\}$, thus $A$ is a reference entities table containing peoples which the names in $A_d$ may represent, e.g. $\{$"John Barnes (computer scientist)", "John Barnes (American author)", "John Barnes (football player)",$\dots\}$. Consider $D = \{d_k|k = 1, \dots,K\}$ is a set of documents containing the names in $A_d$, where possibility $A_t = \{at_l|l = 1,\dots, L\}$ is a set of composition of name tokens (first/middle/last name or in abbreviations), and due to a person has multiple name variations, e.g., "Shahrul Azman Noah (Professor)" has names/aliases as in $\{$"Shahrul Azman Mohd Noah", "Shahrul Azman Noah", "S. A. M. Noah", "Noah, S. A. M.", $\dots\}$. Moreover, the person's name sometimes affected by the background of social communities, like nation, tribes, religion, etc., where a community simply is characterized by the properties in common. For example, some names of Malaysia or Indonesia peoples sometimes insert special terms: "bin", "b", "binti" or "bt" (respect to "son of" or daughter of"), e.g., one name variation of "Shahrul Azman Noah" is "Shahrul Azman b Mohd Noah". In another case, an certain community give the characteristic to community's members, such as the academic community will add the academic degree such as "Prof." (professor) to its members.

Identifying named entity relates to all observed names in $D$, i.e., $A_x = \{ax_o|o = 1, \dots,O\}$, which need to be patterned and disambiguated. The person's names can be rendered differently in online information sources. They are not named with single pattern of tokens, they are not also labeled with unique identifiers, and therefore the names of people also associated with the uncertain things. Text searching relies on matching pattern, searching on a name based on pattern will only match the form a searcher enters in a search box. This causes low recall and negatively affects search precision \cite{branting2002}, when the name of a single person is represented in different ways in the same database \cite{bhattachrya2006}, such as on the motivation above.

Indeed, in scientific publications such as from IEEE, ACM, Springer, etc., all need the
shortened forms of name, especially forenames represented only by initials. However, the shortened form of name is not only makes the name variation, but it creates the name ambiguity in online information sources, such as Web. For example, the name "J. Barnes" can represent "John Barnes (computer scientist)", "Jack Barnes (American communist leader)", "Johnnie Barnes (American football player)", "Johnny Barnes (Bermudian eccentric)", "Joshua Barnes (English scholar)", etc. Name disambiguation is an important problem in information extraction about persons, and is one of themes in Semantic Web. Thus, the persons name can be expressed by using different aliases due to multiple reasons as motives: use of abbreviations, different naming conventions, misspelling, pseudonyms in publication or bibliographies (citations), or naming variations over time. Some different real world entities have the same name, or they share some aliases. So, it is also a semantic problem. We conclude that there are two fundamental reasons of name disambiguation semantically for identifying
entities generally, or persons specially, i.e., 
\begin{enumerate}
\item different entities can share the same name (lexical ambiguity), and 
\item a single entity can be designated by multiple names (referential ambiguity).
\end{enumerate}
Formally, these name disambiguation problems have tasks: 
\begin{enumerate}
\item $\forall a \in A$, there is a relation $\xi$ to assign a list of documents $D$ containing a such that $\xi : A \stackrel{M : N}{\longrightarrow} A_d$, $A_d$ is a subset of $A_x$, where $\xi(a)\in A_d$, 
\item $\forall a \in A$, there is a relation $\zeta : A \stackrel{M : L}{\longrightarrow} A_t$, $A_t$ is a subset of $A_x$, where $\zeta(a) \in A_t$.
\end{enumerate}
Semantically, extracting the keyword from web snippet is to tie $\xi$ and $\zeta$ into a bundle whereby a group of documents exactly associate with one entity only.

\subsection{Proposed Approach}
We start this approach with describing some concepts: 
\begin{enumerate}
\item A word $w$ is the basic unit of discrete data, defined to be an item from a vocabulary indexed by $\{1,\dots,K\}$, where $w_k = 1$ if $k \in K$, and $w_k = 0$ otherwise; 
\item A term $t_x$ consists of at least one word or a sequence of words, or $t_x = (w_1,\dots, w_l)$, $l = k$, $k$ is a number of parameters representing word, and $|t_x| = k$ is size of $t_x$; 
\item Let a web page denoted by $\omega$ and a set of web pages indexed by search engine be containing pairs of term and web page. Let $t_x$ is a search term and a web page contain $t_x$ is $\omega_x$, we obtain $\Omega_x = \{(t_x,\omega_x)\}$, $\Omega_x$ is a subset of $\Omega$, or $t_x \in \omega_x$ in $\Omega_x$. $|\Omega_x| = |\{t_x,\omega_x\}|$ is cardinality of $x$; 
\item Let $t_x$ is a search term. $S = \{w_a,\dots, w_{max}\}$ is a web snippet (briefly snippet) about $t_x$ that returned by search engine, where $max = \pm 50$ words. $L = \{S_i|i = 1,\dots,N\}$ is a list of snippet.
\end{enumerate}

\begin{table}
\centering
\caption{Statistics of our dataset}
\begin{tabular}{llr}\hline
\multicolumn{1}{c}{Personal Name} &
\multicolumn{1}{c}{Position} &
 Number of pages\cr\hline
Abdul Razak hamdan        & Professor &85\cr
Abdulah Mohd Zin          & Professor &90\cr
Shahrul Azman Mohd Noah   & Professor &134\cr
Tengku Mohd Tengku Sembok & Professor &189\cr
Md Jan Nordin             & Professor &41\cr\hline
\end{tabular}
\end{table}

Based on these concepts, we develop an approach based on the overlap principle \cite{nasution2010a} to extract keywords from web snippet. The interpretation of overlap principle by using the query as a composition of $t_x \cap t_y$ or "$t_x$,$t_y$" is to get a reflection the real world from the web, while to implement it we use one of similarity measures, for example, the similarity based on Kolmogorov complexity \cite{nasution2010b}
\begin{equation}
sim(t_x,t_y) = \frac{\log(2|\Omega_x\cap\Omega_y|}{\log(|\Omega_x|+|\Omega_y|)}
\end{equation}
We assume that ambiguity is caused by overlapping interpretations and understanding of the things that exist in the real world. This assumption describes the usefulness of intersection of regular sets. Therefore, we can formulate the conditions of the overlap principle as follows:\\
\begin{itemize}
\item[] For first condition (Condition 1): Let $t_x$ be a term and $t_a$ is a representation term of person name $a \in A$. We define a condition of overlap principle of $t_x$ and $t_a$, i.e. $t_a \cap t_x = \emptyset$, but $t_a, t_x \in \omega_a$ and $t_a,t_x \in \omega_x$, such that $t_a, t_x \in S$.\\ 
\item[] Second condition (Condition 2): Let $t_a,t_x,t_y \in S$ with $|\Omega_x| = | \Omega_y|$. We define a condition of overlap principle between $t_a$ and $t_x$
or ty, i.e. $|\Omega_a \cap \Omega_x| > |\Omega_a \cap\Omega_y|$.
\end{itemize}

\section{Results and Discussion}
\subsection{Evaluation and Dataset}
Consider a set $L$ of documents or snippets, each containing a reference to a person. Let $P = \{P_1,\dots, P_{|P|}\}$ be a partition of $\xi$ and $\zeta$ into references to the same person, so for example $P_i = \{S_1, S_4, S_5, S_9\}$ might be a set of references to "Abdul Razak Hamdan" the information technology professor. Let $C = \{C_1,\dots, C_{|C|}\}$ be a collection of disjoint subset of $L$ created by algorithm and manually validated such that each $S_i$ has a identifier, i.e., URL address, Table 1. Then, we will denote by $L_C$ the references that have been clusters based on collection. Based on measure were introduced \cite{lloyd2005}, we define a notation of recall $Rec()$ as follows: $Rec(S_i) = (|\{S \in P(S_i) : C(S) = C(S_i)\}|)/(|\{S \in P(S_i)\}|)$ and a notation of precision $Prec()$ as follows: $Prec(S_i) = (|\{S \in C(Si) : P(S) = P(S_i)\}|)/(|\{S \in C(S_i)\}|)$ where $P(S_i)$ as a set $P_i$ containing reference $S_i$ and $C(S_i)$ to be the set $C_i$ containing $S_i$. Thus, the precision of a reference to "Abdul Razak Hamdan" is the fraction of references in the same cluster that are also to "Abdul Razak Hamdan". We obtain average recall ($REC$), precision ($PREC$), and $F$-measure for the clustering $C$ as follows:
\begin{equation}
REC = \frac{\sum_{S\in L_C} Rec(S)}{|L_C|}
\end{equation}
\begin{equation}
PREC = \frac{\sum_{S\in L_C} Prec(S)}{|L_C|}
\end{equation}
\begin{equation}
F = \frac{2\cdot REC\cdot PREC}{REC+PREC}
\end{equation}

\begin{table}
\centering
\caption{Recall, Precision, F-measure for "Abdul Razak Hamdan"}
\begin{tabular}{lrrr}\hline
\multicolumn{1}{c}{Keywords} & Recall & Precicion & F-measure\cr\hline
science    & 0.46 & 0.10 & 0.16\cr
Malaysia   & 0.40 & 0.20 & 0.27\cr
data       & 0.38 & 0.10 & 0.16\cr
based      & 0.25 & 0.26 & 0.25\cr
technology & 0.25 & 0.28 & 0.26\cr
study      & 0.24 & 0.30 & 0.27\cr
computer   & 0.23 & 0.40 & 0.29\cr
using      & 0.20 & 0.40 & 0.27\cr
nor        & 0.15 & 0.40 & 0.21\cr
system     & 0.10 & 0.40 & 0.16\cr\hline
Overlap for (2) (3) \& (4) & 0.82 & 0.23 & 0.36\cr\hline
\end{tabular}
\end{table}
\subsection{Experiment}
Let us consider information context of actors, that includes all relevant relationships as well as interaction history, where Yahoo! Search engines fall short of utilizing any specific information, especially context information, and just use full text index search in web snippets. In experiment, we use maximum of $1,000$ web snippets for search term ta representing an actor. The web snippets generate the words list for each actor, outputs very rare words because of the diversity its vocabularies. For example, the list of words for a named actor "Abdul Razak Hamdan" generated a list of $26$ candidate words. For example, we have as many as 85 web pages about Abdul Razak Hamdan in our dataset. By using Yahoo! Search, we execute the query "Abdul Razak Hamdan" and a keyword "science", then we get 39 web pages that contain the name "Abdul Razak Hamdan" and a word "science", and in accordance with our dataset. We obtain this value when loading $387$-th web page, and this value persists until the maximum number of loading of web pages, i.e. 500. Thus, we obtain Rec("Abdulah Razak Hamdan,science") = 46\% and Prec("Abdullah Razak Hamdan,science") = 10\%. There are $11$ words, $\{$science, Malaysia, software, data, based, technology, study, computer, using, nor, system$\}$, and we have the counting of $REC = 0.82$, $PREC = 0.23$ and $F = 0.36$, see Table 2.

\section{Conclusion}

The approach based on overlap principle has the potential to be incorporated into existing method for extracting personal feature like as keyword. It shows how to uncover a keyword by exploiting web snippets and hit counts. Our near future work is to compare some methods for looking into the possibility of enhancing this method.

\end{document}